\documentclass[times, twocolumn]{aastex63}
\usepackage[T1]{fontenc}

\usepackage[hang,flushmargin]{footmisc}
\usepackage{graphicx}
\usepackage{amsmath, amssymb, bm}
\usepackage{upgreek}
\usepackage{epstopdf}

\hypersetup{
    allcolors = {blue},
}

\newcommand{\DMunits}{{\rm pc~cm^{-3}}}
\newcommand{\Swin}{Centre for Astrophysics and Supercomputing, Swinburne University of Technology, P.O. Box 218, Hawthorn, VIC 3122, Australia}
\newcommand{\sydney}{Sydney Institute for Astronomy, School of Physics, University of Sydney, Sydney, NSW 2006, Australia}
\newcommand{\dunlap}{Dunlap Institute for Astronomy and Astrophysics, University of Toronto, 50 St. George Street, Toronto, ON M5S 3H4, Canada}
\newcommand{\curtin}{International Centre for Radio Astronomy Research, Curtin University, Bentley, WA 6102, Australia}
\newcommand{\ATNF}{Australia Telescope National Facility, CSIRO Astronomy and Space Science, P.O. Box 76, Epping, NSW 1710, Australia}
\newcommand{\WVU}{Department of Physics and Astronomy, West Virginia University, Morgantown, WV 26506, USA}
\newcommand{\NRL}{Space Science Division, Naval Research Laboratory, Washington, DC 20375, USA}
\newcommand{\berkeley}{Department of Astronomy, University of California Berkeley, Berkeley CA 94720, USA}
\newcommand{\GWCWVU}{Center for Gravitational Waves and Cosmology, West Virginia University, Chestnut Ridge Research Building, Morgantown, WV 26505, USA}
\newcommand{\ozgrav}{ARC Centre of Excellence for Gravitational Wave Discovery (OzGrav), Australia}

\begin{document}

\title{Faint Repetitions from a Bright Fast Radio Burst Source}
\shorttitle{Detection of repetitions from FRB\,171019}
\shortauthors{Kumar et al.}

\correspondingauthor{Pravir Kumar}
\email{pravirkumar@swin.edu.au}

\author[0000-0003-1913-3092]{Pravir Kumar}
\affiliation{\Swin}

\author[0000-0002-7285-6348]{R. M. Shannon}
\affiliation{\Swin}

\author[0000-0003-0289-0732]{Stefan Os{\l}owski}
\affiliation{\Swin}

\author[0000-0002-9586-7904]{Hao Qiu}
\affiliation{\sydney}
\affiliation{\ATNF}

\author[0000-0003-3460-506X]{Shivani Bhandari}
\affiliation{\ATNF}

\author[0000-0002-0161-7243]{Wael Farah}
\affiliation{\Swin}

\author[0000-0002-4796-745X]{Chris Flynn}
\affiliation{\Swin}

\author[0000-0002-0893-4073]{Matthew Kerr}
\affiliation{\NRL}

\author[0000-0003-1301-966X]{D.R. Lorimer}
\affiliation{\WVU}
\affiliation{\GWCWVU}

\author[0000-0001-6763-8234]{J.-P. Macquart}
\affiliation{\curtin}

\author[0000-0002-3616-5160]{Cherry Ng}
\affiliation{\dunlap}

\author{C.~J. Phillips}
\affiliation{\ATNF}

\author[0000-0003-2783-1608]{Danny C. Price}
\affiliation{\Swin}
\affiliation{\berkeley}

\author[0000-0002-6730-3298]{Ren\'ee Spiewak}
\affiliation{\Swin}
\affiliation{\ozgrav}

\begin{abstract}
We report the detection of repeat bursts from the source of FRB\,171019, one of the brightest fast radio bursts (FRBs) detected in the Australian Square Kilometre Array Pathfinder (ASKAP) fly's eye survey. Two bursts from the source were detected with the Green Bank Telescope in observations centered at 820~MHz. The repetitions are a factor of $\sim$590 fainter than the ASKAP-discovered burst. All three bursts from this source show no evidence of scattering and have consistent pulse widths. The pulse spectra show modulation that could be evidence for either steep spectra or patchy emission.  The two repetitions were the only ones found in an observing campaign for this FRB totaling $1000$ hr, which also included ASKAP and the 64-m Parkes radio telescope,  over a range of frequencies (720--2000\,MHz) at epochs spanning two years. The inferred scaling of repetition rate with fluence of this source agrees with the other repeating source, FRB\,121102. The detection of faint pulses from FRB\,171019 shows that at least some FRBs selected from bright samples will repeat if follow-up observations are conducted with more sensitive telescopes. 
\end{abstract}

\keywords{Radio transient sources (2008), Transient sources (1851), Fast radio bursts}

\section{Introduction}\label{sec:intro}
We are now starting to unravel the enigmatic astrophysical phenomenon of fast radio bursts (FRBs), millisecond-duration transient events first discovered over a decade ago \citep{lorimerburst}. The observed dispersion measures (DMs) of FRBs significantly exceed the expected contribution from the Milky Way \citep{thornton_frbs}, suggesting extragalactic origins. The localization of several bursts sources (\citealt{r1_localization_chatterjee, askap_localized, ravi_localization}) unequivocally places them at cosmological distances; nevertheless, their physical origin has yet to be determined.

\begin{deluxetable*}{ccccccccr} 
\tabletypesize{\small} 
\tablecolumns{8} 
\tablewidth{0pt} 
\tablecaption{Details of FRB\,171019 follow-up observations \label{tab:followupobs}} 
\tablehead{ 
\colhead{Telescope} & \colhead{Receiver} & \colhead{Gain}  & \colhead{T$_\mathrm{sys}$} & \colhead{Central Frequency} & \colhead{Bandwidth} & \colhead{Beam FWHM} & \colhead{Sensitivity\tablenotemark{a}} & \colhead{Obs. Time} \\ 
\colhead{}          & \colhead{}         & \colhead{(K Jy$^{-1}$)}& 
\colhead{(K)}             & \colhead{(MHz)}              & \colhead{(MHz)}    & \colhead{(\arcmin)} & \colhead{(Jy ms)}       & \colhead{(hr)}}
\startdata 
ASKAP  &    PAF          & 0.1 & 50 & 1297.5 & 336 & 60 & 51.8 & 986.6 \\
Parkes & Multibeam       & 0.7 & 23 & 1382 & 340 & 14  & 1.10 & 12.4  \\
GBT    & Prime Focus 1   & 2.0 & 20 &  820 & 200 & 15  & 0.27 & 9.7\\
GBT    & L-band          & 2.0 & 20 & 1500 & 800 & 9 & 0.13 & 0.9
\enddata 
\tablenotetext{a}{The limiting fluence for a pulse width of 5 ms and S/N threshold of 7.5$\sigma$ for GBT, 9.5$\sigma$ for ASKAP and 10$\sigma$ for Parkes, as discussed in Section 2.}
\end{deluxetable*}

There are currently about 100 FRB sources published \citep[FRBCAT\footnote{\url{http://www.frbcat.org}; visited 2019 August 26.};][]{frbcat}, most of which have only been detected once. The repeat bursts from FRB\,121102 \citep{repeater_1} enabled precise localization of the burst source and the identification of its host galaxy \citep{r1_localization_chatterjee, r1_localization_tendulkar}. The existence of repetitions ruled out cataclysmic progenitor scenarios for the origin of its emission.
Since its discovery, more than 100 bursts (\citealt{zhang_frbs_r1, Hessels19}) have been detected from this source in a broad range of frequencies, from as high as 8 GHz \citep{r1_highestfreq_gajjar} to as low as 600 MHz \citep{r1_lowestfreq_josephy}. 
The discovery of second repeating source, FRB\,180814 \citep{repeater_2}, with properties similar to FRB\,121102, strengthened evidence for the existence of a substantial population of repeating FRB sources. Recently the Canadian Hydrogen Intensity Mapping Experiment (CHIME) telescope reported detection of eight new repeating FRB sources \citep{chime_mega_repeaters}.

The localization of the ostensibly one-off (single pulse detection, which has not been shown to repeat) FRB\,180924 to a position 4 kiloparsecs from the center of a luminous galaxy at a redshift of $z=0.32$ \citep{askap_localized}, enabled the first comparison of burst host galaxies. 
The massive  ($\sim 10^{10} M_\odot$) host galaxy of FRB~180924 is in stark contrast with the low-mass ($\sim 10^8 M_\odot$), low-metallicity dwarf galaxy of the repeating source FRB\,121102 \citep{r1_localization_tendulkar}, thus raising questions whether there are multiple FRB formation channels.  Recently, another burst (FRB\,190523) has also been localized to $10''\times2''$  uncertainty, and associated with a massive ($\sim 10^{11} M_\odot$) host galaxy  \citep{ravi_localization}, partially based on the agreement between the burst DM ($760.8$\,$\DMunits$) and the galaxy redshift ($z=0.66$).   
%FRB 181112 has been localised to a $10^{9.4} M_\odot$ galaxy at a redshift of $z=0.48$.

One of the most exciting open questions is the relationship between the repeating and one-off FRB sources. It is not clear whether all FRBs repeat. Are there two (or more) classes of FRBs, or are the one-off FRBs simply the most energetic bursts from repeating sources? The absence of repeat bursts even after hundreds of hours of follow-up (\citealt{ravi_carina, ravi_2016}) and the diversity in properties (e.g., temporal structure and polarization) of one-off FRBs could be evidence for multiple populations of FRBs (\citealt{caleb_repeating, james_limits}). However, in a recent analysis, \cite{ravi_repeating} has suggested that the volumetric rate of one-off FRBs is inconsistent with the rate of all possible cataclysmic FRB progenitors and concludes that most FRBs are repeating sources.

Among the strongest constraints on FRB repetition so far come from \cite{askap_nature} with the discovery of 20 FRBs in the first Commensal Real-time ASKAP Fast Transient
\citep[CRAFT\footnote{\url{https://astronomy.curtin.edu.au/research/craft/}};][]{craft_start} survey. The survey was conducted using a ``fly's eye'' configuration to maximize sky coverage at a Galactic latitude of $|b| = 50 \pm 5~\deg$ and a central frequency of 1.3 GHz. The survey produced a well-sampled population of FRBs and established a relationship between burst dispersion and observed luminosity. The mean spectral index for these bursts ($\alpha \approx -1.5$, where $E_{\nu} \propto \nu^\alpha$) is found to be similar to that of the normal pulsar population \citep{jp_askapfrbs}. A key feature of the survey was that it revisited the same positions hundreds of times over its duration, producing $ \sim$ 12,000\,hr (\citealt{askap_nature,craft_performance_james}) of (self) follow-up observations, which included times before and after bursts were detected.
No repeat bursts from detected FRBs were found in the survey. 

One possible reason for the lack of repeat detections is that ASKAP is insufficiently sensitive to faint repetitions from the bursts. Conducting follow-up observations with more sensitive instruments will be more effective \citep{cordes_chatterjee}; for example, Parkes has a repeat detection rate $\sim 10^4$ times greater than ASKAP, assuming the luminosity distribution follows a power-law where, above some luminosity $\mathcal{L}$, the number of detections \(N (> \mathcal{L}) \propto \mathcal{L}^{\alpha}\) assuming $\alpha = -2$ \citep{connor_petroff_repetitions}. To complement the ASKAP self follow-up, we have also been conducting sensitive monitoring campaigns of ASKAP detections with the 64-m Parkes radio telescope and the 110-m Robert C. Byrd Green Bank Telescope (GBT). 
The arcminute localization of FRBs, made possible by the multi-beam detection \citep{bannister_askap} using ASKAP's phased-array feed (PAF) enabled the follow-up of FRB fields with large aperture telescopes.

\begin{figure*}[t]
	\centering
\includegraphics[scale=0.7]{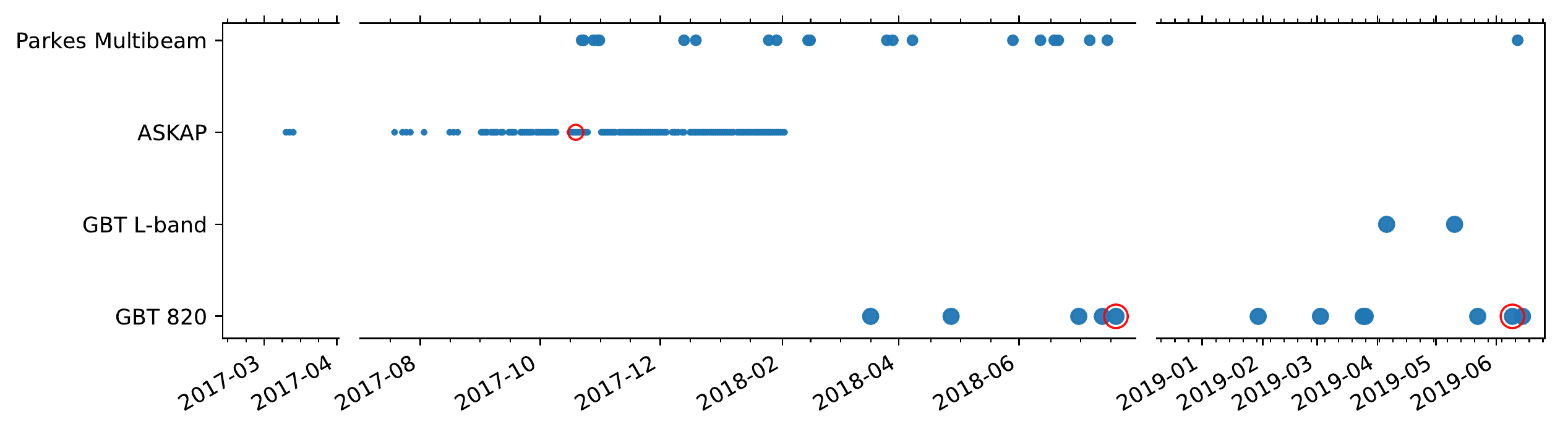} 
\figcaption{Timeline of follow-up observations of FRB\,171019. Each row represents a set of observations from a given radio telescope. Observations with bursts are encircled with red. The first repeat burst is found in be observation dated 2018 July 20 and the second one on 2019 June 9. }
\label{fig:timeline}
\end{figure*}

In this Letter, we report\footnote{Analysis of the entire campaign is ongoing and  will be reported elsewhere.}  the discovery of repetitions from FRB\,171019, one of the brightest bursts found in the ASKAP fly's-eye survey. The burst was $\sim 5$\,ms wide with a measured fluence of $220$~Jy\,ms \citep{askap_nature}. The observed DM was 460 $\DMunits $, a factor of 11 in excess of the NE2001 model \citep{cordes2001model} prediction along that line of sight. In Section \ref{sec:observations}, we describe the observational campaigns for this FRB. In Section \ref{sec:repeats}, we present the properties of the repeat pulses. In Section \ref{sec:discussion}, we discuss the implications for the FRB population as a whole.

\section{Observations and Data Processing}\label{sec:observations}

We searched for repeat pulses from FRB\,171019 using ASKAP, Parkes, and the GBT. The observing details for all three telescopes used are summarized in Table \ref{tab:followupobs}. Each telescope was pointed at the position of FRB\,171019 reported in \cite{askap_nature}, i.e., R.A. = $22^{\rm{h}}17^{\rm{m}}32^{\rm{s}}$ and decl. = $-08\arcdeg39\arcmin32\arcsec$ (J2000.0 epoch). This position was obtained with $10'\times10'$ uncertainty (90\% confidence) as described in \cite{bannister_askap}. As such, the positional uncertainty was well within the full-width at half maximum (FWHM) of the follow-up telescopes. Figure \ref{fig:timeline} shows a timeline of the radio observations of FRB\,171019. 

\subsection{ASKAP  Searches}
ASKAP follow-up was conducted in fly's eye configuration with each antenna pointing at a different position in the sky, and the survey regularly revisiting the same positions \citep{askap_nature}. FRB searches are performed in near-real-time using  {\tt FREDDA} \citep{fredda_ascl}, a GPU-based implementation of the fast dispersion measure transform algorithm \citep[FDMT;][]{fdmt}. For a description of the detection methods and search pipeline, see \cite{bannister_askap}. We found no other astrophysical events at similar DMs of FRB\,171019 exceeding a threshold signal-to-noise ratio (S/N) of 9.5 (which corresponds to a fluence sensitivity of 52 Jy ms for a pulse duration of 5 ms) in 987 hr of observations.

\subsection{Parkes  Searches}
At Parkes, we used the 20-cm multibeam receiver to search for bursts from FRB\,171019, using the Berkeley-Parkes Swinburne Recorder (BPSR) mode of the HI-Pulsar system to record full-stokes spectra with 64 $\upmu$s time and 390\,kHz frequency resolution \citep{StaveleySmith:1996, Price:2016}. 
The search process \citep{ppta_oslowski} was similar to that of the SUrvey for Pulsars and Extragalactic Radio Bursts project ``Fast'' pipeline (SUPERB; details in \citealt{parkes_multibeam_keith, parkes_multibeam_keane}). 
The online pipeline stored the 8-bit data stream from all 13 beams in a ring buffer over the bandwidth of 340 MHz centered at 1382 MHz. The data were then searched using {\tt Heimdall} \citep{barsdell12} up to a maximum DM of 4096 $\DMunits$ with a tolerance (S/N loss tolerance between each DM trial) of 20 \%. The transient pipeline sorts candidate FRB events from radio interference using the methods detailed in \cite{parkes_multibeam_bhandari}. The pipeline searched for bursts above a threshold S/N of 10, thus sensitive up to a fluence of $1.1$ Jy\,ms for a burst of width similar to FRB\,171019.
No bursts were found in all the 12.4 hr of observations at the dispersion measure of FRB\,171019. 

\begin{figure*}[!ht]
\begin{center}
\begin{tabular}{c@{\hskip 0.1in}c@{\hskip 0.1in}c}
  \includegraphics[scale=0.30]{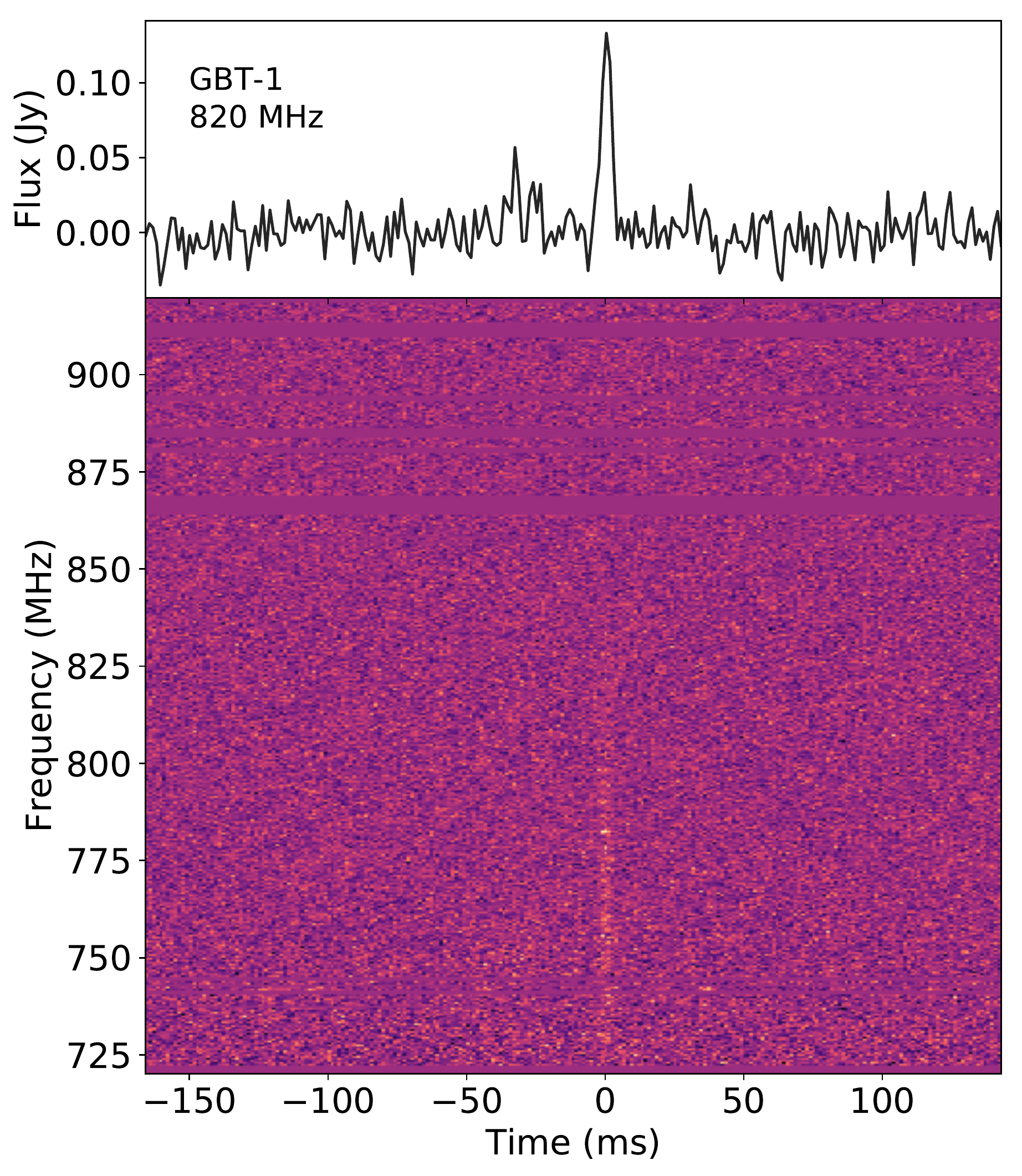} &  
  \includegraphics[scale=0.30]{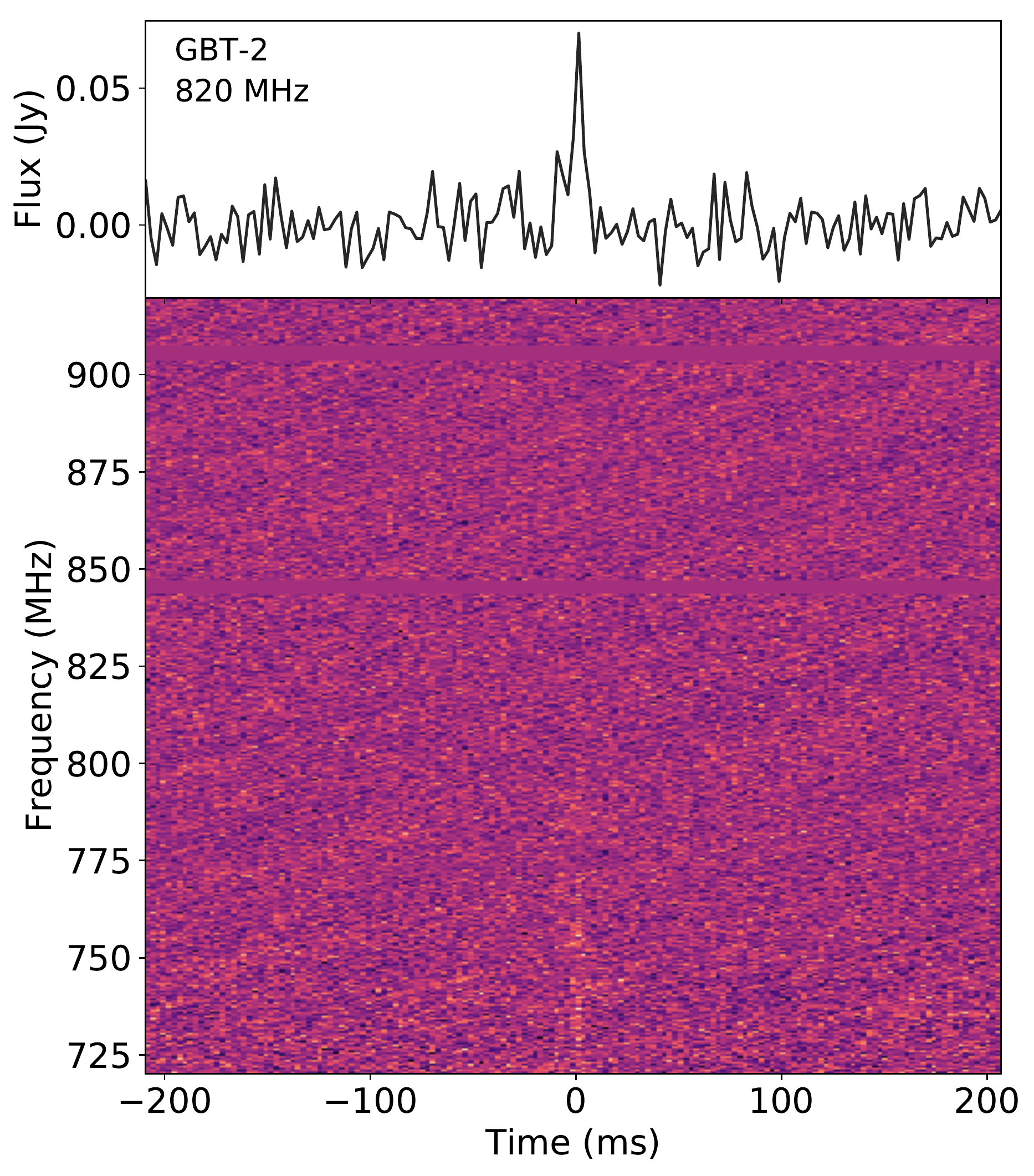} &  
  \includegraphics[scale=0.30]{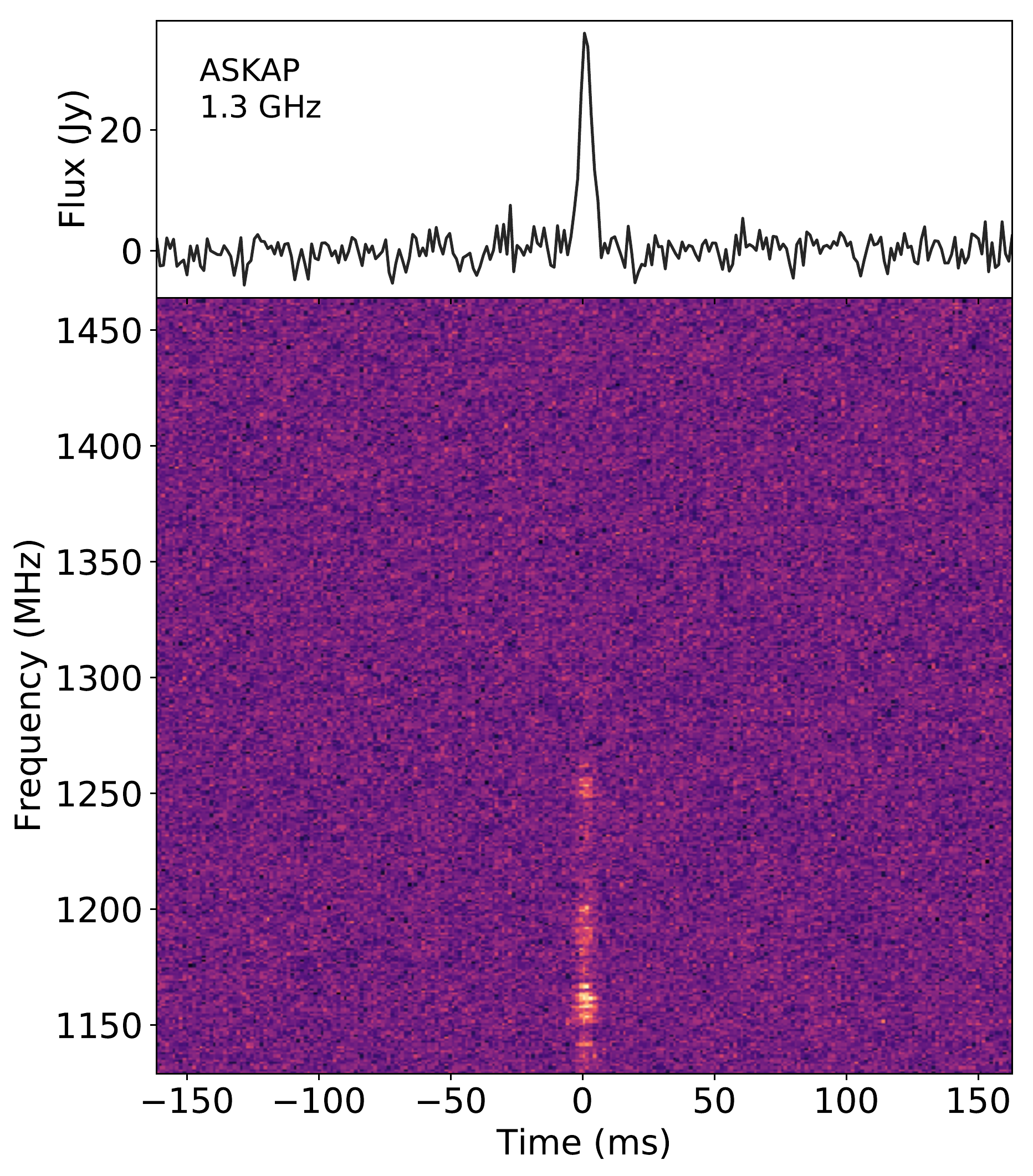} \\
\end{tabular}
\end{center}
\caption{
Dynamic spectra for both repeat bursts detected at GBT and ASKAP FRB\,171019 dedispersed at their optimal DM. From the left: repeat burst 1 (resolution = 1.31 ms), repeat burst 2 (resolution = 2.62 ms), and ASKAP FRB\,171019 (resolution = 1.26 ms). For each burst, the top panel shows the flux density averaged over frequency channels.
\label{fig:repeaterplots}  }
\end{figure*}

\begin{deluxetable*}{cccccccc}
\tabletypesize{\small} 
\tablecolumns{8} 
\tablewidth{0pt} 
\tablecaption{Properties of detected bursts. Bursts properties calculated for full bandwidth appear in numbered rows and for lower half band in row next to them in chronological order. \label{tab:burstsproperties}}
\tablehead{ 
\colhead{No.}        & \colhead{Telescope} & \colhead{TOA\tablenotemark{a}}      & \colhead{Fluence\tablenotemark{b}} & \colhead{Gaussian FWHM} & \colhead{Integrated} & \colhead{Spectral Index\tablenotemark{d}} & \colhead{DM\tablenotemark{e}}\\ 
\colhead{}           &                    & \colhead{(MJD)}      & \colhead{(Jy ms)} & \colhead{(ms)}          & \colhead{S/N\tablenotemark{c}} &   & ($\DMunits$)   }
\startdata 
0 & ASKAP & 58045.56061371(2) & 219 $\pm$ 5 & 5.4 $\pm$ 0.3 & 24.8 & $-$12.6 $\pm$ 1.4 & 461 $\pm$ 1\\  
  &       &  & 388 $\pm$ 12    & 5.2 $\pm$ 0.2 & 32.4 & $-$9.9 $\pm$ 2.0 & \\
1 & GBT   & 58319.356770492(1) & 0.60 $\pm$ 0.04 & 4.0 $\pm$ 0.3 & 15.2 & $-$7.8 $\pm$ 1.2 & 456.1 $\pm$ 0.4\\ 
  &       &  & 1.11 $\pm$ 0.07 & 4.2 $\pm$ 0.3 & 16.7 & $-$0.9 $\pm$ 1.8  &\\
2 & GBT   & 58643.321088777(1) & 0.37 $\pm$ 0.05 & 5.2 $\pm$ 0.8 &  7.9 & $-$13.2 $\pm$ 2.8  & 457 $\pm$ 1\\ 
  &       &  & 0.61 $\pm$ 0.07 & 3.7 $\pm$ 0.5 &  9.1 & $-$9.6 $\pm$ 3.3  &\\
\enddata 
\tablenotetext{a}{Burst time of arrival is referenced at the highest frequency (1464 MHz  for ASKAP and 920 MHz for GBT). The ASKAP burst arrival time is measured in TAI, while GBT burst arrival times are in UTC. Uncertainties are in parentheses.}
\tablenotetext{b}{SEFD curve of GBT-820 MHz is taken from \url{ https://science.nrao.edu/facilities/gbt/proposing/GBTpg.pdf}. Fluence error ranges correspond to an uncertainty of one in S/N. For ASKAP burst, fluence is taken from \cite{askap_nature}.} 
\tablenotetext{c}{S/N is the signal-to noise ratio calculated with width of the pulse as twice the Gaussian FWHM.}
\tablenotetext{d}{These are forced fits based on the assumption of a power-law spectrum. For GBT-1 spectrum, the fit obtained in the frequency range (820, 750 Mhz) of the lower half band is $-$6.0 $\pm$ 2.8. }
\tablenotetext{e}{DM for ASKAP burst has been corrected from the value in \cite{askap_nature} to account for an identified  1~MHz offset in frequency labeling.}
\end{deluxetable*}

\subsection{GBT  Searches}
The GBT observations were obtained with the Prime Focus 1 (centered at 820\,MHz) and L-band receivers (details in Table \ref{tab:followupobs}), and data recorded with the Green Bank Ultimate Pulsar Processing Instrument \citep[GUPPI;][]{2008SPIE.7019E..1DD}.  Each pointing was sampled with a time resolution of 81.92 $\upmu$s and 2048 frequency channels (512 channels for the L-band receiver), and written to a PSRFITS format file with full-Stokes parameters.

To search the GBT data for bursts, we first converted the PSRFITS data to total intensity SIGPROC\footnote{\url{http://sigproc.sourceforge.net}} filterbank format. The dynamic spectra were then normalized to remove the receiver bandpass by scaling each channel to a mean of zero and standard deviation of unity. Using the PRESTO\footnote{\url{https://github.com/scottransom/presto}} \citep{ransom_presto} tool {\tt rfifind} and the median absolute deviation statistics, we identified bad channels affected by radio frequency interference (RFI). The resulting data were then searched using {\tt Heimdall} for dispersed pulses. We performed two searches: a narrow search within the DM range of 446 to 474 $\DMunits$ over 220 trials using a tolerance of 1\% and then a wider search in a DM range of 0 to 2000 $\DMunits$ with a tolerance of 5\%. Candidates satisfying the following criteria were retained for further analysis: S/N $\geq 6.5 $ (7.5 for the wider search), pulse width $\leq 41.94 $ ms and members\footnote{Number of individual boxcar/DM trials clustered into a candidate.} $\geq 2 $. For the L-band data, we also apply a minimum threshold for pulse width (0.65\,ms)  to mitigate false-positives produced by spurious narrow-band short-duration candidates. 

\begin{figure*}[!ht]
\begin{center}
\begin{tabular}{ccc}
\includegraphics[scale=0.29]{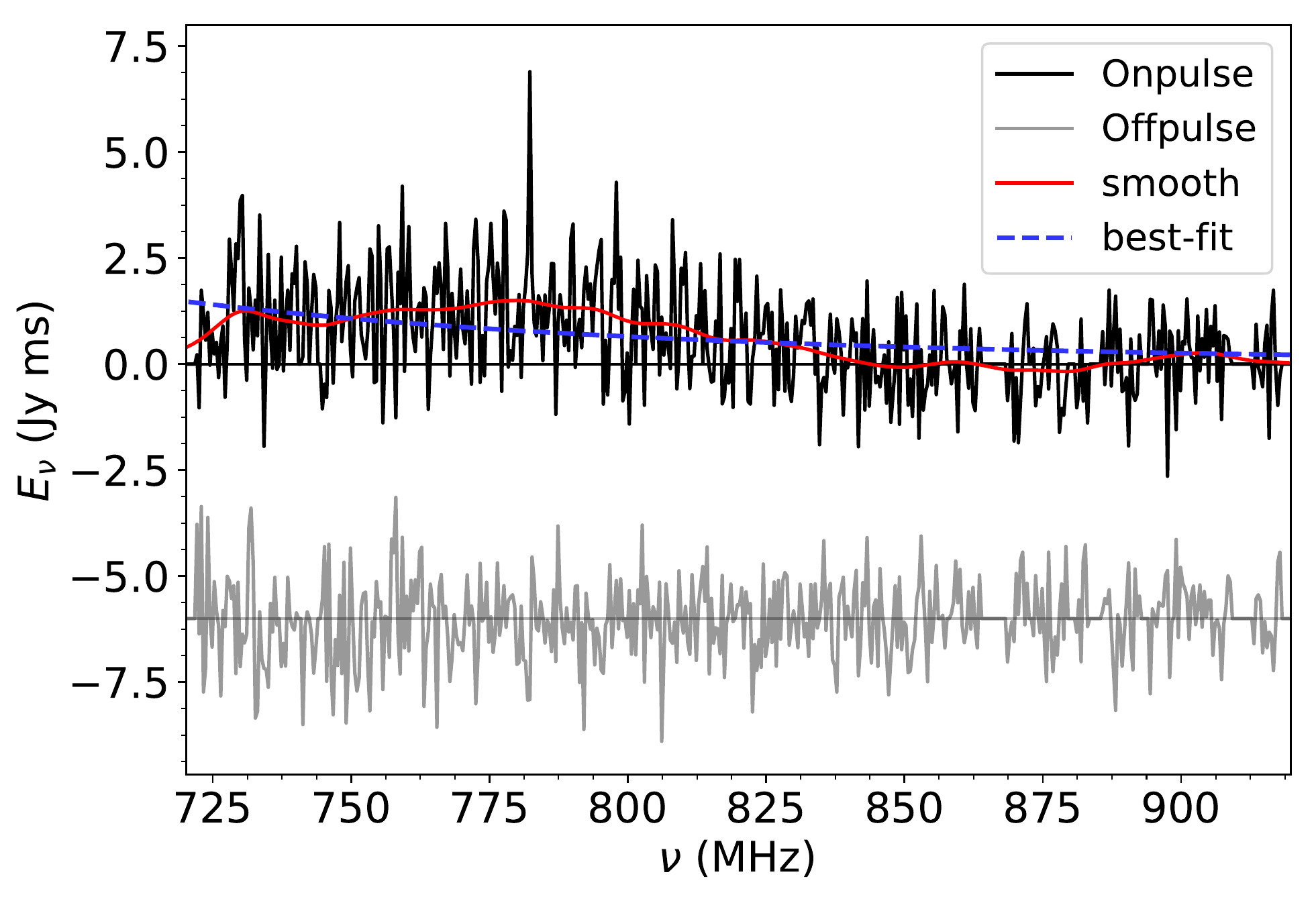} & \includegraphics[scale=0.29]{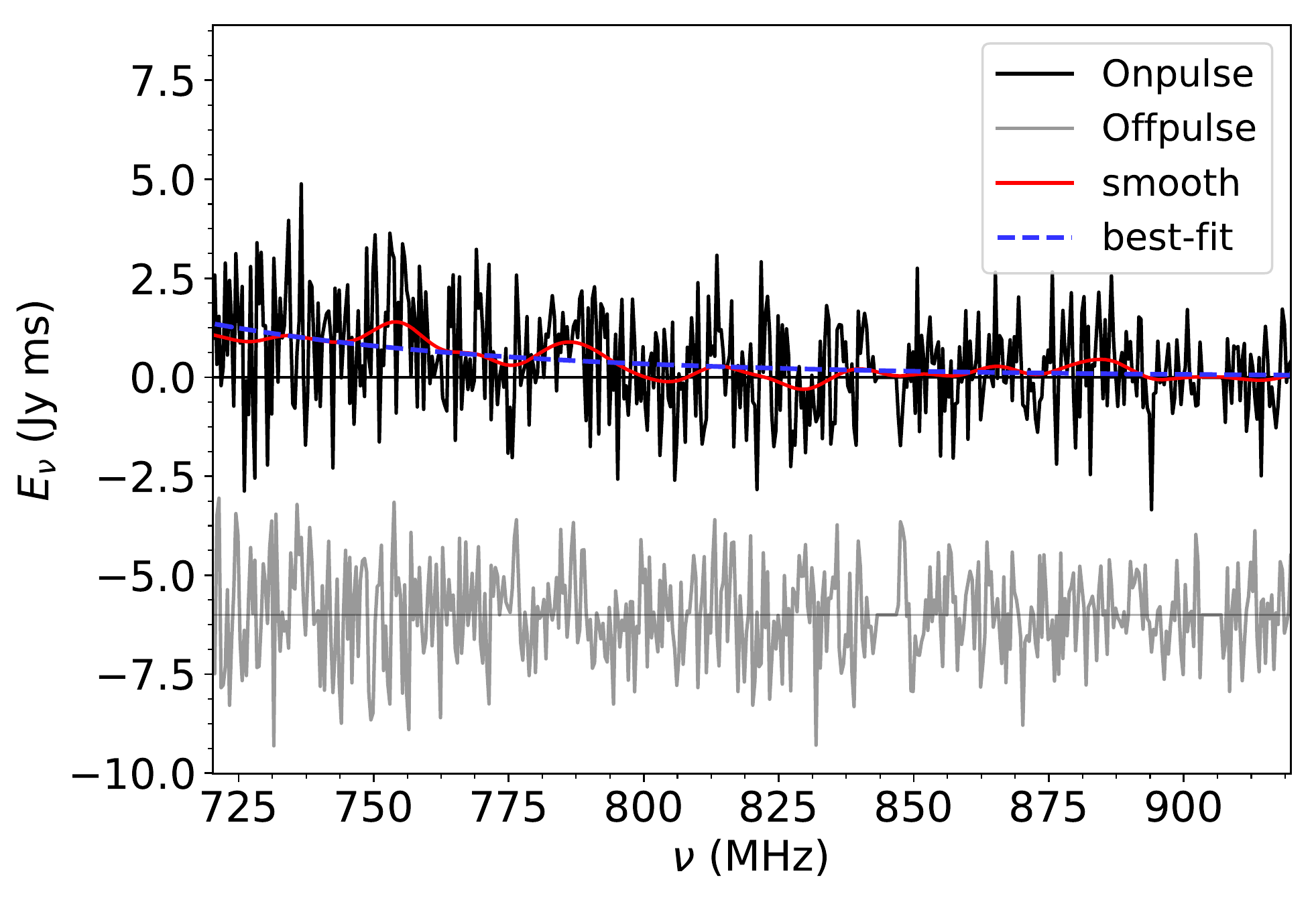} & 
\includegraphics[scale=0.29]{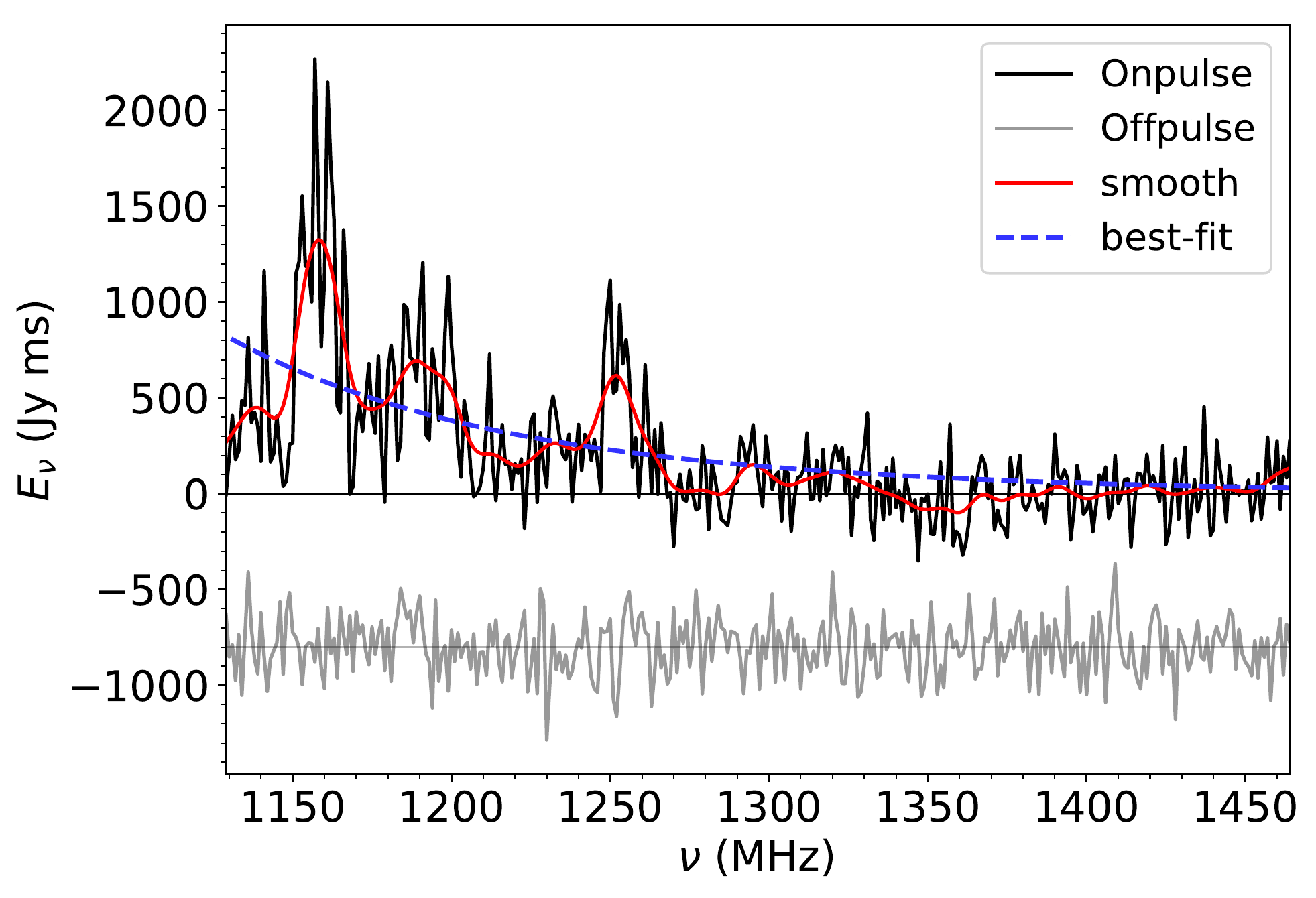} \\
\includegraphics[scale=0.29]{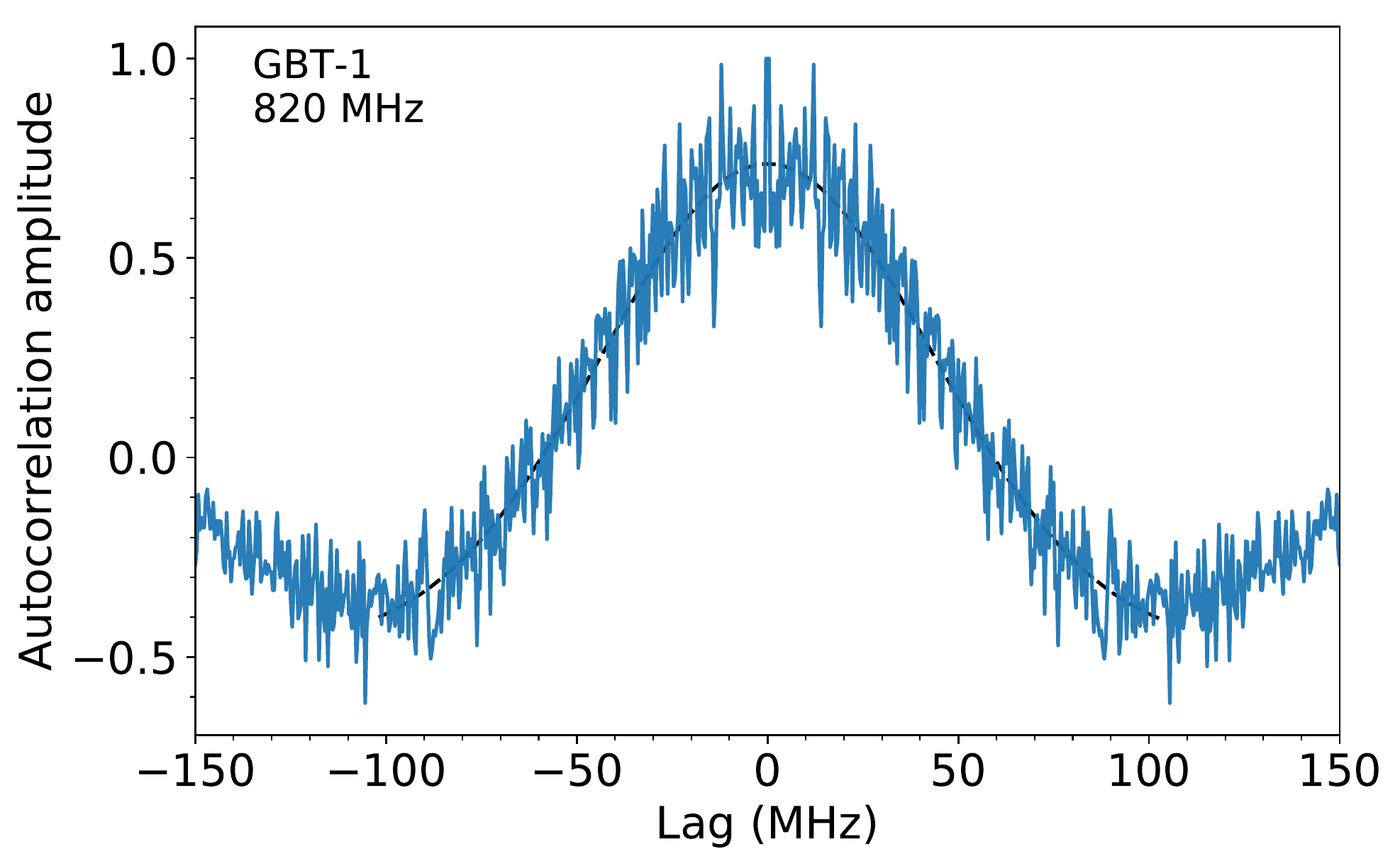} & \includegraphics[scale=0.29]{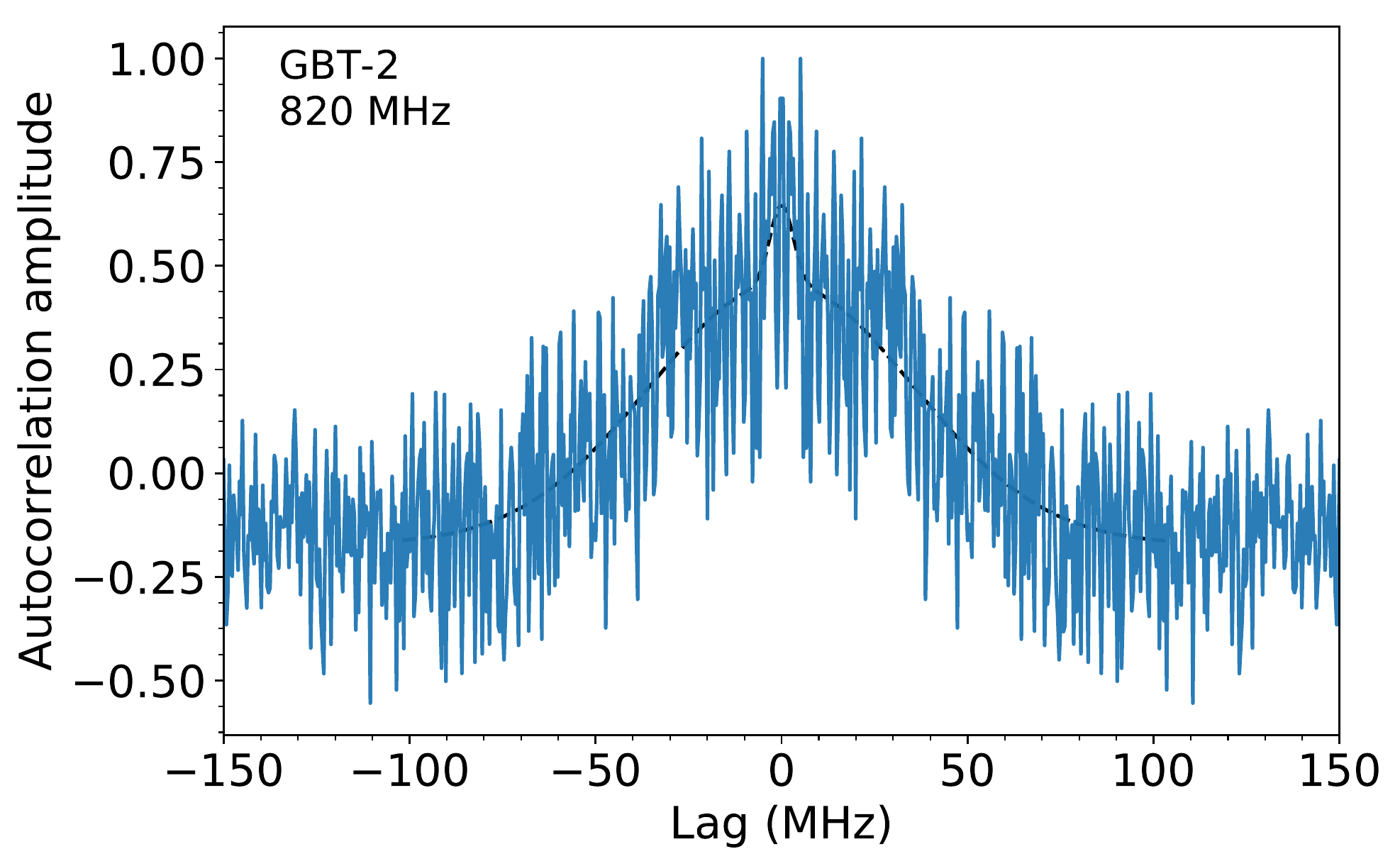} & 
\includegraphics[scale=0.29]{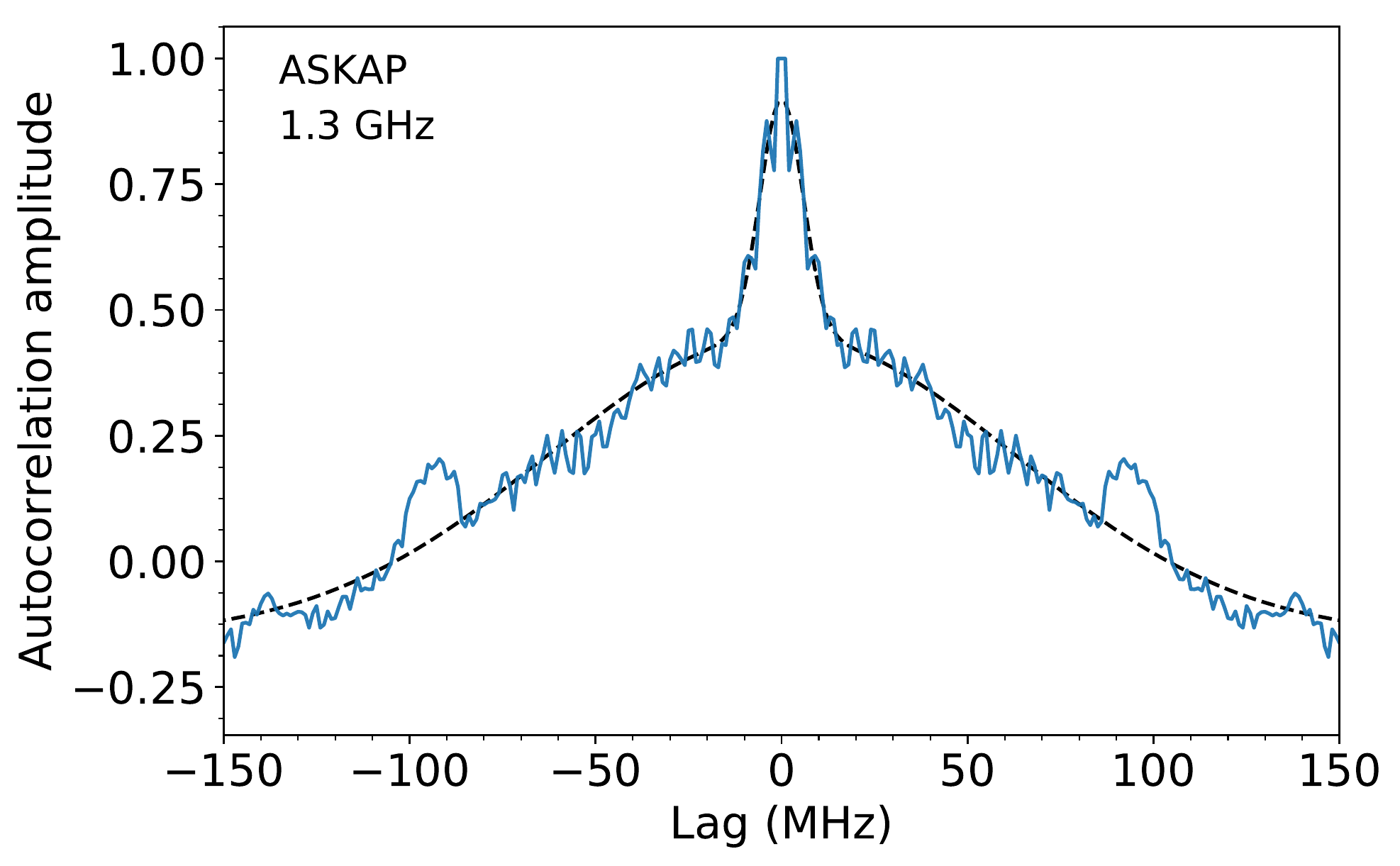} 
\end{tabular}
\end{center}
\caption{ Burst spectra and autocorrelation functions. From left: GBT-1 ($\Delta \nu $ = 0.39 MHz), GBT-2 ($\Delta \nu $ = 0.39 MHz), ASKAP detection ($\Delta \nu $ = 1 MHz). {\em Upper panels:}  burst spectra. Red lines are smoothed spectra (using a Gaussian kernel with standard deviation of 4 MHz). Blue dashed lines are best-fit power law model. Gray lines are off-pulse baseline spectra, and are offset from zero for clarity. Horizontal lines show zero power for both the on- and off-pulse spectra.
{\em Bottom panels}:  autocorrelation function of the time-averaged spectrum of bursts. The zero lag value, which is associated with self noise present in spectrum, has been removed.} 
\label{fig:spectrabursts}
\end{figure*}

We used deep neural network trained models, as developed by \cite{fetch}\footnote{All 11 trained models are taken from \url{https://github.com/devanshkv/fetch}} to perform the FRB/RFI binary classification of the candidates. Following their prescription, we created dedispersed frequency-time and DM-time image data for each candidate, which were then classified using {\tt keras} \citep{keras} with the {\tt TensorFlow} \citep{tenserflow} backend. We took the union of all the 11 model predictions and visually inspected each one of the resulting FRB candidates to identify astrophysical pulses. We found two bursts (hereafter GBT-1 and GBT-2) at similar DM to that of FRB\,171019 in the observations.

\subsection{GBT Periodicity Searches}
We also conducted a search for periodicity in the GBT data using Fourier domain searching with the PRESTO routine {\tt accelsearch}, as  well as time domain searching using the Fast Folding Algorithm (FFA\footnote{Based on \url{https://bitbucket.org/vmorello/riptide}}) package {\tt riptide}. Before searching, frequency channels and time blocks significantly affected by RFI were identified using {\tt rfifind} and masked. The data were corrected for dispersion over 240 trial DMs evenly spaced from 400 to 520 $\DMunits$, generating a time series at each trial. We used {\tt dedisp} \citep{barsdell12}, a GPU-accelerated package, to create time series. The FFA-based periodicity search was carried out to find long-period signals, where we searched periods ranging from 0.2 to 10\,s. We detected no significant periodic astrophysical signal in the data above a S/N threshold of 10 (chosen to minimize the number of false-positive candidates).

\section{The Repeat bursts}\label{sec:repeats}
The two repeat bursts were detected in 820 MHz GBT observations 9 and 20 months after the initial ASKAP detection, and are marked with red circles in Figure \ref{fig:timeline}. The dynamic spectra of the bursts are shown in Figure \ref{fig:repeaterplots}, along with the original detection at ASKAP.
To measure the width of the bursts, we fit the frequency-averaged pulse profile with a Gaussian model and report the FWHM\footnote{The measured FWHM values are consistent with the W50 estimates (width at 50\% of pulse peak).} ; both bursts are approximately 4.5\,ms in duration. The residual after subtracting the best-fit model from pulse profiles appears to be white, thus there is no underlying temporal sub-structure in the dynamic spectrum of either repeat burst. The burst durations are well in excess of the maximum DM smearing across a channel for the GBT data, which is 1.0 ms. For reference, we also calculate the properties of the ASKAP detection. The time resolution for ASKAP data is 1.26 ms with a maximum DM smearing of 2.66 ms present within a channel. All three bursts are visible in the lower half of the band but not detected in the top half. Thus, the lower sub-band fluences are larger than the full-band averaged values. The burst properties obtained from the full band as well as from the lower half of the band are listed in Table  \ref{tab:burstsproperties}. The spectral structures of the bursts are described in Section \ref{sec:substructure}.

\subsection{Scattering and dispersion analysis}

To obtain scattering timescales and burst DMs, we perform multi sub-band modeling of the burst pulse profiles using the nested sampling method  \texttt{Dynesty} \citep{dynesty} implemented in the parameter estimation code \texttt{Bilby} \citep{bilby}. 
We model each of the pulse profiles to be a Gaussian convolved with an exponential pulse-broadening function. 
The broadening time $\tau$ is assumed to vary with frequency with a fixed index,  $\tau \propto \nu^{-4}$. 
We model both interchannel dispersion delay (which causes the pulse to arrive at different times in different sub-bands) and intrachannel dispersion smearing (which increases the pulse width in quadrature with an intrinsic width).
Based on the ratio of Bayesian evidence between models with and without scattering, we conclude that the data do not support presence of scattering.
For the ASKAP pulse, we limit the scattering timescale to be $ < 3.52$ ms, at a reference frequency of  1\,GHz. For the repeat bursts, we group the lower half band of the data into four sub-bands to perform the analysis; we limit the scattering time scales (referenced to 1~GHz) to be $< 0.79$ ms and $< 1.77$ ms for GBT-1 and GBT-2 respectively. In contrast, the optimized DMs of the bursts shown in Table  \ref{tab:burstsproperties} suggest that the repetitions have a different apparent DM than the higher-frequency ASKAP detection.

\subsection{Polarization Properties}
We extracted the GBT/GUPPI data for the detected repeat bursts using \texttt{dspsr} \citep{dspsr} producing a full-Stokes archive file.
We found no evidence for linear or circular polarization in the pulse data. 
It is possible that the non-detection of linear polarization is the result of Faraday rotation of the burst through magnetized plasma. 
We searched for Faraday rotation using the PSRCHIVE (\citealt{psrchive, psrchive2}) \texttt{rmfit} routine in the range ${\rm |RM|} \leq 3 \times 10^4 \,{\rm rad\, m^{-2}}$ (this is the rotation measure (RM) at which the polarization position angle rotates by one radian in one frequency channel at the center of the band), but no significant RM was found.  We note that no polarization calibration procedures were conducted during GBT observations. For the ASKAP burst, only the total intensity data were retained; hence, no polarimetric properties could be derived from this burst.

\subsection{Spectral Properties} \label{sec:substructure}
The spectrum for each burst shown in Figure \ref{fig:spectrabursts} is formed by integrating the signal over the time samples within twice the measured FWHM of the frequency-averaged pulse. The amplitude of each spectrum was then scaled to fluence, using the system equivalent flux density (SEFD) and the radiometer equation.  Modest changes to the window do not significantly affect estimates of fluence. 
All three bursts show lower fluences at higher frequencies. One possibility is that the bursts have steep spectra. We characterize this by fitting a power-law model $ E_{\nu} \propto \nu^{\alpha} $. Spectral indices, $ \alpha $ obtained from the fits to individual spectra are in Table \ref{tab:burstsproperties}. All three bursts show steep spectra in the observed bands with $ \alpha $ ranging from $-$13 to $-$8. While both the ASKAP burst and the GBT-2 spectra is extremely steep in the lower half band as well, the GBT-1 spectrum is nearly flat.

Off-axis attenuation is unlikely to significantly change the fluences or spectral indices of the repetitions. Based on the posterior distribution from the ASKAP multi-beam localization in \cite{askap_nature}, the median correction to the fluence results in an increase of 8\%, and is $<  24\%$ with 90\% confidence. The median spectral index correction is $-0.07$, and with 95\% confidence is less than $< - 0.2$. This analysis assumes the GBT beam can be modeled as a Gaussian with an FWHM (beamwidth) of $15\arcmin$ at 820 MHz (Table \ref{tab:followupobs}). We therefore rule out any primary beam offset as the cause of the observed steep spectra for the GBT pulses.

The spectral modulation in the bursts could be intrinsic to the emission or due to the propagation effects. To characterize this, we calculate the autocorrelation function (ACF) of the burst spectra \citep{wael_frb} as shown in Figure \ref{fig:spectrabursts}. We fit the ACF with Gaussian component models using a non-linear optimization approach \citep{lmfit} to find the frequency scales of characteristic modulation in spectra. We detect two characteristic frequency scales in the ASKAP spectrum of band extent 13 and 147 MHz. For the GBT-1 spectrum, the ACF can be best described with a single component (100 MHz), which is the total bandwidth over which the pulse is visible. We observe a bright spike in the spectrum (at $\sim 776$ MHz), but its width is comparable to the channel width. It is unclear if this is astrophysical or RFI. For the GBT-2, apart from the frequency scale of 82 MHz, we also see marginal evidence for a second component (7 MHz wide). However, because the second component is not present in an analysis of the lower half of the band where the burst is bright, it is most likely due to RFI or noise fluctuations. We also estimate the amplitude of the spectra variability using the square of the modulation index $ m^2$, by computing the mean-normalized spectral autocovariance \citep{jp_askapfrbs} from the spectrum of bursts. The estimated values of $ m^2$ for the three bursts are 2.4, 1.1, and 1.9 respectively.

\begin{figure}[!ht]
\begin{center}
\begin{tabular}{ccc}
\includegraphics[scale=0.6]{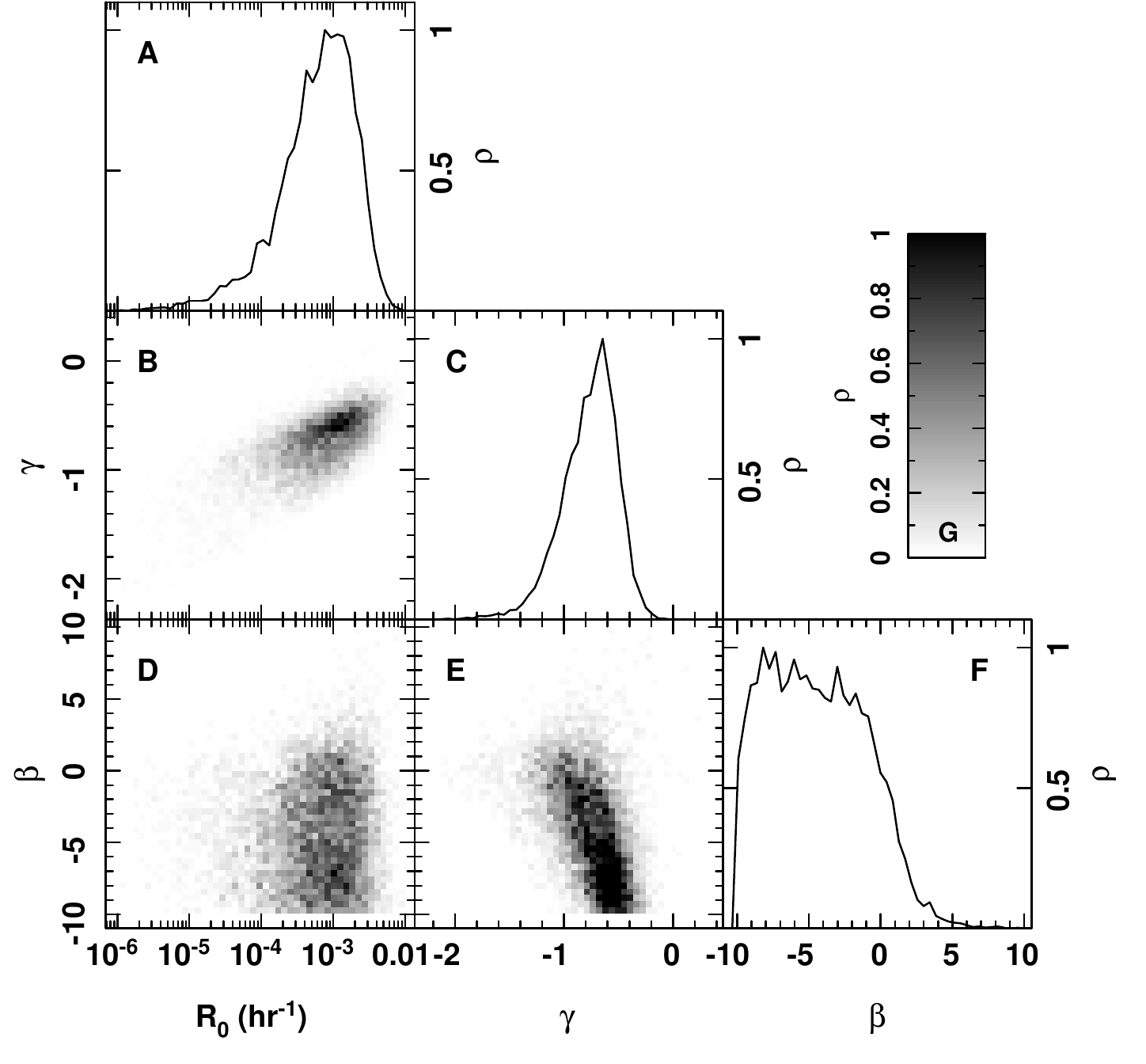} & 
\end{tabular}
\end{center}
\caption{Posterior distributions for burst rate parameters.  Panels A, C, and F show the one-dimensional marginalized distributions for $R_0$, $\gamma$, and $\beta$, with peak probability densities normalized to unity. 
Panels B, D, and E show the two-dimensional distributions (normalized again such that peak probability density are unity), with grayscale shown in panel G. The rate $R_0$ has been scaled to ASKAP sensitivities and frequencies ($52$\,Jy\,ms; see Table \ref{tab:followupobs}). } \label{fig:rate_corner}
\end{figure}

\subsection{Inferring the repetition rate}
We use Bayesian methodology to characterize the repetition statistics of FRB\,171019,  given the detection of pulses with ASKAP at 1.3 GHz and the GBT at 820 MHz, and the non-detections with the GBT at 1.5 GHz and Parkes at 1.3 GHz. 
We assume that the cumulative burst rate above a fiducial fluence $S$ at a frequency $\nu$ is  
\begin{equation}
R(>S, \nu) = R_0 \left(\frac{S}{S_0(\nu)}   \right)^\gamma,
\end{equation}
where $R_0$ is the rate of bursts above fluence 
\begin{equation}
S_0(\nu) = S_0  \left(\frac{\nu}{\nu_0}\right)^{\beta}
\end{equation}
at a frequency $\nu_0$.

We assume that the burst event rate in a survey $i$ of total integration time $T_i$ will follow Poisson statistics with rate parameter $\lambda_i = T_i R(>S_i, \nu_i)$, where $\nu_i$ is the observing frequency of the survey and $S_i$ is the survey sensitivity.
In this case we can infer the parameters in the survey $R_0$,  $\beta$, and $\gamma$ using the likelihood 
\begin{equation}
L = \prod_{i=1}^{N_s} \frac{1}{n_i!}  e^{- \lambda_i}  \left( \lambda_i \right)^{n_i},
\end{equation}
where $n_i$ is the number of bursts found in survey $i=1$ to $N_s$. 

We  sample the posterior distribution using the \texttt{multinest} algorithm \citep{2009MNRAS.398.1601F} assuming uniform priors on $\beta$ and $\gamma$ ($-10 < \beta, \gamma < 10$), and logarithmic priors on $R_0$ between $10^{-6}$ and $1$\,hr$^{-1}$, where the reference frequency $\nu_0=1.3$~GHz and sensitivity $S_0=52$\,Jy\,ms. 
We do not take into account the spectral index obtained for bursts (Table \ref{tab:burstsproperties}) in this repetition analysis, which allows for an independent estimation of the spectral index. 
The posterior distribution is shown in Figure \ref{fig:rate_corner}.  
We find that the slope of the burst intensity distribution is consistent with a power-law distribution with an index between $-1.5 \lesssim \gamma \lesssim 0$.  The value depends strongly on the spectral dependence of the burst emission rate $\beta$.
The inferred steep values of $\beta$ ($\beta  \ll  -1.5 $,  with the lower prior acceptable) are consistent with the observed spectra (in the case, the spectrum is attributed to a steep power-law process),  but inconsistent with the ASKAP population overall \citep{jp_askapfrbs}.  
The observed shallow values of $\gamma$ are consistent with observations of the first repeating FRB\,121102 \citep{law_repeating}.

\section{Discussion and  Conclusions}\label{sec:discussion}

The bursts in FRB\,171019 extend over the range of 219 Jy\,ms to 0.37 Jy\,ms, a fluence range of $\sim$ 590. 
At a similar frequency range, this is a factor of $\sim$ 3 larger than what has been observed in FRB\,121102 (\citealt{gourdji_lower_limit, Hessels19}) and an order of magnitude larger than any other repeating FRB source \citep{chime_mega_repeaters}. The wide range in observed fluences shows that, like Galactic pulsars and magnetars, repeating FRB sources can emit pulses with a wide range of luminosities, and that repeating sources can emit bright pulses like the initial ASKAP detection. The inferred isotropic peak luminosity of bursts ranges from $L\sim6\times 10^{43}$ erg s$^{-1}$ to $L\sim6\times 10^{40}$ erg s$^{-1}$, nearly 3 orders in magnitude. Models for burst emission need to account for this wide range.

We find evidence for variations in the apparent DMs of the pulses. It is unclear whether the difference is genuine DM variation or due to non-dispersive effects as has been observed in FRB\,121102 \citep{Hessels19}. We note that this discrepancy in apparent DM can also be due to the different volumes of the medium being probed by the ASKAP and the GBT. All three bursts are temporally resolved with similar widths. We note that the pulse width of the GBT-2 is less reliable when measured in the whole band due to the presence of RFI in the upper half of the band. However, taking the DM smearing and sampling time into account, the intrinsic width of all bursts are consistent within uncertainties. We find no evidence for sub-structure in the pulse profile as seen in other FRBs (\citealt{wael_frb, Hessels19}). We would be insensitive to any sub-structure narrower than $\sim$ 1 ms in GBT detections.

The band extent of spectral features differs between ASKAP and GBT pulses and is inconsistent with diffractive scintillation. The burst exhibited a large degree of spectral modulation in the original ASKAP detection. It was not clear whether the bright structures were intrinsic to the burst or due to propagation effects \citep{jp_askapfrbs}. If the spectral structures observed in the ASKAP detection were the result of diffractive scintillation, we would expect the band extent of the structures present in the GBT pulses to be factor of  $(\nu_{\rm GBT}/\nu_{\rm ASKAP})^{-4} \approx 6 $ smaller. The widest structures in the ASKAP burst (width approximately half the band) would be observed to be $\sim 25$\,MHz wide in the GBT spectrum. However, we only see evidence for structures much wider than this in the GBT observations. We do not find any conclusive evidence of diffractive scintillation in repeat bursts.

All of the bursts from FRB\,171019 are only visible in lower half of their respective bands, which could be evidence of an extremely steep spectrum. This argument is also consistent with the non-detection of repetitions at Parkes and GBT L-band receivers. If we assume this steep spectrum ($ \sim -9$) to be the case, it provides a very natural way to understand the detection of repetitions from this source in the context of all the non-detections (\citealt{askap_nature, james_limits}) from other ASKAP FRBs (assuming a non-negligible fraction are repeaters). 
It would make the repeat bursts at least a factor of $(\nu_{\rm ASKAP}/\nu_{\rm GBT})^{9} \approx 60 $ fainter at the center frequency of ASKAP. In that scenario, the fluence discrepancy between ASKAP and the GBT detection is actually $ > 10^4 $, assuming a constant spectral index that makes FRB \,171019 special within the ASKAP population of flatter-spectrum FRBs \citep{jp_askapfrbs}. However, we are cautious not to over-interpret this result, as there are not many physical mechanisms to produce such a steep spectrum. 
It is quite possible that the spectrum is similar to patchy emission, seen in the other repeater FRB sources (\citealt{michilli_repeater1, chime_mega_repeaters}). This is evident as the GBT-1 spectrum is nearly flat in the lower half of the band. Also, the ASKAP detection has a large spectral modulation that can not be explained by scintillation. In this scenario, the power-law model might not be the correct approach for the spectral index measurement \citep{mwa_steep}. 

Another possibility is FRBs having stochastic patchy or modulated emission in different parts of the frequency band for an individual burst but, when ensemble-averaged, produce steep spectra as observed in ASKAP one-off FRBs sample (\citealt{askap_nature, jp_askapfrbs}). This would be tested with further detections.

The other published repeating burst sources (\citealt{repeater_2, Hessels19}) share common features such as spectra variability, sub-structures in their dynamic spectrum, and sub-components in pulse profile. We do not observe any of these features in all three bursts. A coherently dedispersed detection from FRB\,171019 with high time resolution will provide more information on these distinctions. 
The bursts from the FRB\,171019 source are fainter at higher frequencies, which is not the case with many of the bursts from  the  FRB\,121102 source, where bursts have been reported brighter at higher frequencies \citep{gourdji_lower_limit}.
The first detection of FRB\,171019 comes from a different sample of bright FRBs \citep{askap_nature} than the CHIME detections \citep{chime_frbs} and FRB 121102.
The host galaxy of a localized burst \citep[FRB\,180924;][]{askap_localized} from the ASKAP population also originates from a galaxy significantly different to that of FRB\,121102. It will be interesting to see if all repeating FRBs have similar environments as of FRB\,121102. If not, it could be indicative of a different channel for producing repeat burst sources. The detection of further repetitions from this source\footnote{We note that CHIME has recently detected a repeat burst from this source \citep{chime_atel}.}  and localization to a host galaxy will be key to understanding the nature of FRB\,171019 and its relation to other repeating burst sources.

\acknowledgments

We thank C. James for useful discussions. 
P.K., R.M.S., S.O., and R.S. acknowledge support through Australian Research Council (ARC) grant FL150100148. 
R.M.S. and J.P.M. acknowledge support through ARC grant DP180100857.
R.M.S. also acknowledges support through ARC grant CE170100004.
D.R.L. was supported by the National Science Foundation through the award number OIA-1458952.
This work was performed on the OzSTAR national facility at Swinburne University of Technology. OzSTAR is funded by Swinburne University of Technology and the National Collaborative Research Infrastructure Strategy (NCRIS).
Work at NRL is supported by NASA. 
The Green Bank Observatory is a facility of the National Science Foundation operated under cooperative agreement by Associated Universities, Inc.
The Parkes radio telescope  is part of the Australia Telescope National Facility which is funded by the Australian Government for operation as a National Facility managed by CSIRO.
The Australian SKA Pathfinder is part of the Australia Telescope National Facility which is managed by CSIRO. Operation of ASKAP is funded by the Australian Government with support from the National Collaborative Research Infrastructure Strategy. ASKAP uses the resources of the Pawsey Supercomputing Centre. Establishment of ASKAP, the Murchison Radio-astronomy Observatory and the Pawsey Supercomputing Centre are initiatives of the Australian Government, with support from the Government of Western Australia and the Science and Industry Endowment Fund. We acknowledge the Wajarri Yamatji people as the traditional owners of the Observatory site.

\bibliography{references}{}
\bibliographystyle{aasjournal_revised}

\end{document}